\documentclass[reprint, amsmath, amssymb, aps, superscriptaddress, prb]{revtex4-2}

\usepackage{graphicx}
\usepackage{dcolumn}% Align table columns on decimal point
\usepackage{bm}% bold math
\usepackage[colorlinks=true, urlcolor=black, linkcolor=black, citecolor=black]{hyperref}
\usepackage{amsbsy}
\usepackage{lipsum}
\usepackage{verbatim} 
\usepackage{mathtools,amsmath}
\usepackage[version=4]{mhchem}
\usepackage[dvipsnames]{xcolor}
\usepackage{longtable}
\usepackage{multirow}

\begin{document}

\title{Anomalous conductance steps in three-dimensional topological insulator HgTe-based quantum point contacts}

 \author{Elisabeth\,Richter}
 \email{Corresponding authors: elisabeth.richter@ur.de,\\ dieter.weiss@ur.de}
\affiliation{Institut für Experimentelle und Angewandte Physik, Universität Regensburg, 93053 Regensburg, Germany}
 \author{Michael\,Barth}
\affiliation{Institut für Theoretische Physik, Universität Regensburg, 93053 Regensburg, Germany}
 \author{Dmitriy\,A.\,Kozlov}
\affiliation{Institut für Experimentelle und Angewandte Physik, Universität Regensburg, 93053 Regensburg, Germany}
\affiliation{Rzhanov Institute of Semiconductor Physics, 630090 Novosibirsk, Russian Federation}
 \author{Angelika\,Knothe}
\affiliation{Institut für Theoretische Physik, Universität Regensburg, 93053 Regensburg, Germany}
 \author{Nikolay\,N.\,Mikhailov}
\affiliation{Rzhanov Institute of Semiconductor Physics, 630090 Novosibirsk, Russian Federation}
 \author{Juliane\,Steidl}
\affiliation{Institut für Experimentelle und Angewandte Physik, Universität Regensburg, 93053 Regensburg, Germany}
 \author{Cosimo\,Gorini}
\affiliation{Institut für Theoretische Physik, Universität Regensburg, 93053 Regensburg, Germany}
\affiliation{SPEC, CEA, CNRS, Université Paris-Saclay, 91191 Gif-sur-Yvette, France}
 \author{Stefan\,Hartl}
\affiliation{Institut für Experimentelle und Angewandte Physik, Universität Regensburg, 93053 Regensburg, Germany}
 \author{Wolfgang\,Himmler}
\affiliation{Institut für Experimentelle und Angewandte Physik, Universität Regensburg, 93053 Regensburg, Germany}
 \author{Klaus\,Richter}
\affiliation{Institut für Theoretische Physik, Universität Regensburg, 93053 Regensburg, Germany}
 \author{Dieter\,Weiss}
 \email{Corresponding authors: elisabeth.richter@ur.de,\\ dieter.weiss@ur.de}
\affiliation{Institut für Experimentelle und Angewandte Physik, Universität Regensburg, 93053 Regensburg, Germany}

\date{\today}

\begin{abstract}
We explore electrical transport through a point contact in strained HgTe, a three-dimensional topological insulator. In the absence of a magnetic field $B$, there is no quantization.
However, under higher magnetic fields, we observe distinct non-integer conductance steps. Based on numerical tight-binding calculations and a phenomenological Landauer-Büttiker approach, we attribute these atypical, non-integer quantized plateaus to significant scattering effects at the point contact.
\end{abstract}

\maketitle

\section{Introduction}
Quantum point contacts (QPCs), i.e. narrow constrictions between two wider regions of an electron system are a fundamental component of mesoscopic physics. In the case of point contacts in a two-dimensional electron system (2DES), a quantized conductance in units of $2e^2/h$ has been observed \cite{vanWees88,Wharam1988}, highlighting the wave nature of the electrons and the energy quantization of the electronic states in the constriction. These experiments confirmed the relationship between conductance and transmission, a concept proposed by Landauer nearly 70 years ago (see Ref. \cite{vanHouten1996} and references therein). Conductance quantization in units of $2e^2/h$ has also been observed in metallic three-dimensional (3D) point contacts (see Ref. \cite{Krans1995,Requist2016} and references therein), although the precision of quantization is less exact than in their semiconductor 2D counterparts. 

In a four-point geometry, the conductance or resistance across the point contact can be described using the Landauer-Büttiker approach, which expresses these quantities in terms of the modes transmitted or reflected at the point contact \cite{Buttiker1986}. Within this picture, the resistance across a ballistic constriction is given by $\frac{r}{Nt}\frac{h}{e^2}$, where $r$ is the reflection probability of an edge channel, $t$ is the transmission probability through the constriction and $N$ is the number of (spin-resolved) modes. In fully ballistic systems or in the quantum Hall regime, $r$, $t$, and $N$ are integers.
Such point contacts made of a 3D topological insulator (TI) have been predicted to be potential probes for Majorana bound states (MBS)  if on one side of the point contact there is a superconducting wire, produced by s-wave proximity \cite{Wimmer2011}.  The presence of MBS would be indicated by an unusual sequence of conductance steps given by $(n+\frac{1}{2})\frac{4e^2}{h}$ with $n=0,1,2,3...$. Similar unusual quantization in the  presence of MBS states has been envisioned in a 1D wire with one normal and one superconducting contact \cite{deJuan2014}.

Here, we explore normal point contacts made of a 3D TI, i.e., strained HgTe. In addition to the usual valence and conduction band carriers, 3D TIs have Dirac-type surface states that form a 2DES  "wrapped" around the bulk of the TI. The surface states are not spin-degenerate and the spin of an electron is locked to its wave vector $k$ \cite{Hasan2010}. Point contacts made in a two-dimensional topological insulator (2D TI) have been explored, demonstrating the typical sequence of conductance steps observed in a normal point contact in a 2DES, except for an anomaly at the lowest conductance step \cite{Strunz_2020}.  Previous work on the conductance of long topological nanowires has demonstrated the existence of robust topological surface states \cite{Ziegler_2018}. However, point contacts in 3D TIs and potential quantization effects have not yet been explored.

\section{Samples and experimental techniques}

\begin{figure}
\centering
\includegraphics[width=1\columnwidth]{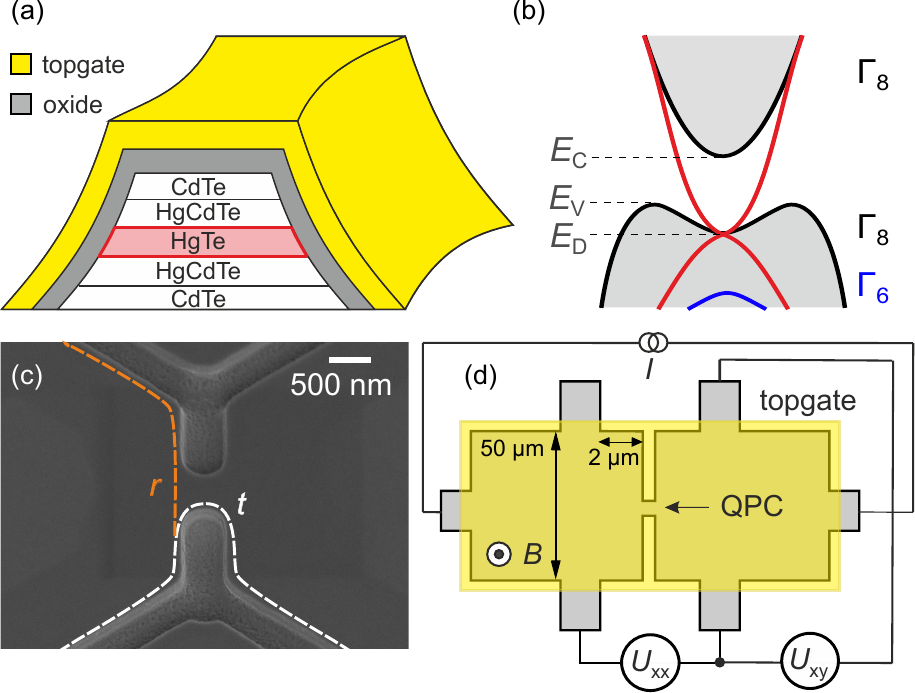}
\caption[] {
(a) Schematic cross section of a QPC (not to scale). The QPC is defined via wet chemical etching in a quantum well structure grown by MBE, consisting of a HgTe layer embedded within CdTe. An HgCdTe buffer layer is incorporated between the two materials to reduce lattice mismatch and enhance carrier mobility. This structure is subsequently covered by an oxide layer (gray) consisting of \ce{SiO2} and \ce{Al2O3} and a metallic topgate (yellow). The topological surface states are highlighted in red.
(b)  Simplified band structure of the three-dimensional topological insulator HgTe, showing surface states (red) within the band gap. The \(\Gamma_8\) (black) and \(\Gamma_6\) (blue) bands, the edges of the conduction and valence bands, as well as the theoretical Dirac point, are indicated. 
(c) SEM image showing a representative HgTe topological insulator (TI) quantum point contact (QPC), at a tilt angle of 50° prior to the deposition of the topgate structure. Wet-chemically etched trenches, about 500 nm deep, delineate the dimensions of the QPC. The image highlights both a transmitted and a reflected edge channel. 
(d) Measurement setup in four-point geometry, where both the longitudinal signal through the quantum point contact (\(xx\)) and the Hall signal (\(xy\)) are measured.
}
\label{Figure1}
\end{figure}

The investigated 3D TI QPCs were fabricated from (013)-oriented, strained 30-, 50- and 80-nm thick HgTe thin films grown by molecular beam epitaxy (MBE) on (013)-oriented GaAs substrates (for details see Ref. \cite{Dima_2016_Quantum_Capacitance}). A cross section of the finished QPC structure is shown in Fig. \ref{Figure1} (a), a scanning electron microscope (SEM) micrograph is shown in Fig. \ref{Figure1} (c).
A lattice mismatch of 0.3\% between HgTe and CdTe, in which the TI layer is embedded, leads to strain in the HgTe thin film and causes a gap to open between the degenerate \(\Gamma_8\) bands, which originate from light hole and heavy hole bands of the trivial CdTe band structure \cite{Dai2008}. This allows access to topological surface states connecting the \(\Gamma_6\) and  \(\Gamma_8\) bands. The size of the transport gap at the  \(\Gamma\)-point is on the order of 15 - 20\,meV \cite{Gospodaric2020, Dima_3D_transport}. The band structure is shown schematically in Fig. \ref{Figure1} (b).
The fabrication process is described in previous work \cite{Ziegler_2018}.

The measurements were conducted using a four-terminal setup [illustrated in Fig. \ref{Figure1} (d)] over a  temperature range from 80\,mK up to 10\,K. 
Both a dilution refrigerator and conventional \ce{^{4}He} cryostats with variable temperature insert were utilized, operating in a magnetic field up to 8\,T. Standard AC lock-in techniques were employed, along with Femto voltage preamplifiers featuring high input resistance of 1\,T$\Omega$. Excitation amplitudes and frequencies typically ranged from 1–10 nA and 1–13 Hz, respectively, effectively mitigating both capacitive coupling and heating effects. To suppress radio frequency noise, \(\pi\)-filters were applied.
A total of 15 devices was tested.

\section{Experimental results}

\begin{figure}
\centering
\includegraphics[width=1\columnwidth]{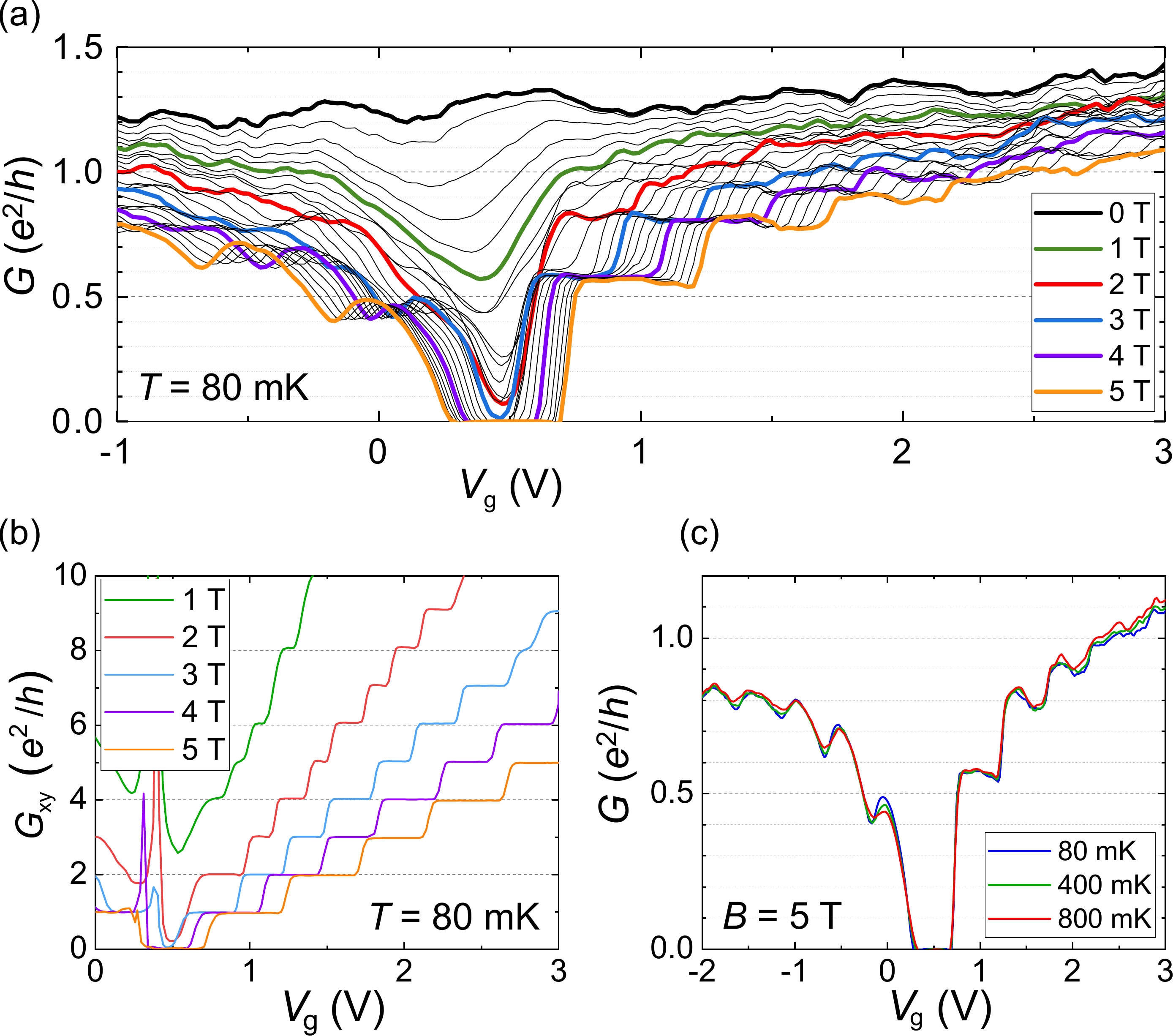}
\caption[] {
(a) The four-terminal QPC conductance $G(V_{\rm g})$ for device A1 as a function of gate voltage is shown for \(B = 0 - 5\)\,T in steps of 200\,mT. The data at 0 T, 1 T, ...5 T are highlighted in different colors. For large magnetic fields, steps emerge at values of 0.6, 0.8 and 1.0\,$e^2/h$. The contribution from the macroscopic regions of the sample is negligible. Data is for device A1.
(b) Hall conductance $G_{\rm xy}(V_{\rm g})$  for \(B\) = 1 - 5\,T in steps of 1\,T at the same temperature shows well-developed QHE states with precise quantization. %Macroscopic resistivity was measured in the sample part  to the side of the QPC.
(c) QPC conductance $G(V_{\rm g})$ for different temperatures between 80\,mK and 800\,mK (for device A1). The insensitivity of the plateau position to temperature indicates the single-electron nature of the phenomenon.
} 
\label{Figure2}
\end{figure}

The results of four-terminal QPC conductance ($G=I/U_{xx}$ ) measurements versus gate voltage $V_{\rm g}$ at $T = 80$\,mK for device A1, representative of all group A devices (group information detailed below), are shown in Fig.~\ref{Figure2} (a). At zero magnetic field, the measured conductance exhibits a value on the order of $e^2/h$, weak dependence  on the gate voltage, absence of plateaus, and presence of random yet  reproducible fluctuations. Evidently, the structure under study is characterized by pronounced scattering, likely due to sharp lithographic boundaries of the constriction or disorder. One method to mitigate backscattering, both in the QPC and its vicinity,  is to apply an out-of-plane magnetic field \cite{Halperin1982}. 
As the applied magnetic field increases in Fig.~\ref{Figure2} (a), 
the measured conductance systematically decreases, while also displaying a clear dependence on the gate voltage. At $B = 3$\,T, when the macroscopic part of the sample has already entered the quantum Hall effect (QHE) mode, distinct steps have developed in the conductance
across the QPC on the electron side. Surprisingly, the steps do not appear at integer values of $e^2/h$, but rather at non-integer values of 0.6\,$e^2/h$ for the first step, 0.8\,$e^2/h$ for the second one etc.  
The presented family of curves, conducted over a wide range of magnetic fields (up to 5\,T) and in increments of 200\,mT, demonstrates that the position of these steps remains consistent from their initial appearance to the maximum magnetic field of 5\,T.  This consistency suggests the presence of a scattering mechanism within the QPC, enabling a fixed fraction of electrons to traverse the device irrespective of the magnetic field strength. While less pronounced, similar steps are observed on the hole side, where the current is primarily carried by valence band holes. However, we focus our analysis on the electron side below. 

To confirm the preservation of the high quality of the 2DES, similar to the observation in Ref. \cite{Ziegler_2020_LL_3D}, we measure the transverse conductance in the macroscopic region of the sample, i.e. away from the QPC.  The data in Fig.~\ref{Figure2} (b) show a well-developed QHE state already at $B=3$\,T with precise $G_{\rm xy}$ quantization and all filling factors resolved, proving the high quality of the sample. We also examine the response of the QPC conductance to temperature changes. The data presented in Fig.~\ref{Figure2} (c) clearly shows that a temperature change from 80 to 800\,mK maintains the plateau position. Measurements on sample A3 up to 10 K (not shown) additionally confirmed the weak temperature dependence of the steps.
%A significant effect of temperature is only observed at $T>5$\,K, when the plateaus begin to smear, and the conductance at the charge neutrality point at about 0.5\,V also increases due to trivial activation. 
The insensitivity of the observed plateaus to temperature changes suggests a lack of connection with the 0.7-anomaly \cite{0.7-anomaly}, implying the probable single-electron nature of the scattering responsible of the anomalous conductance quantization.

\begin{figure}
\centering
\includegraphics[width=1\columnwidth]{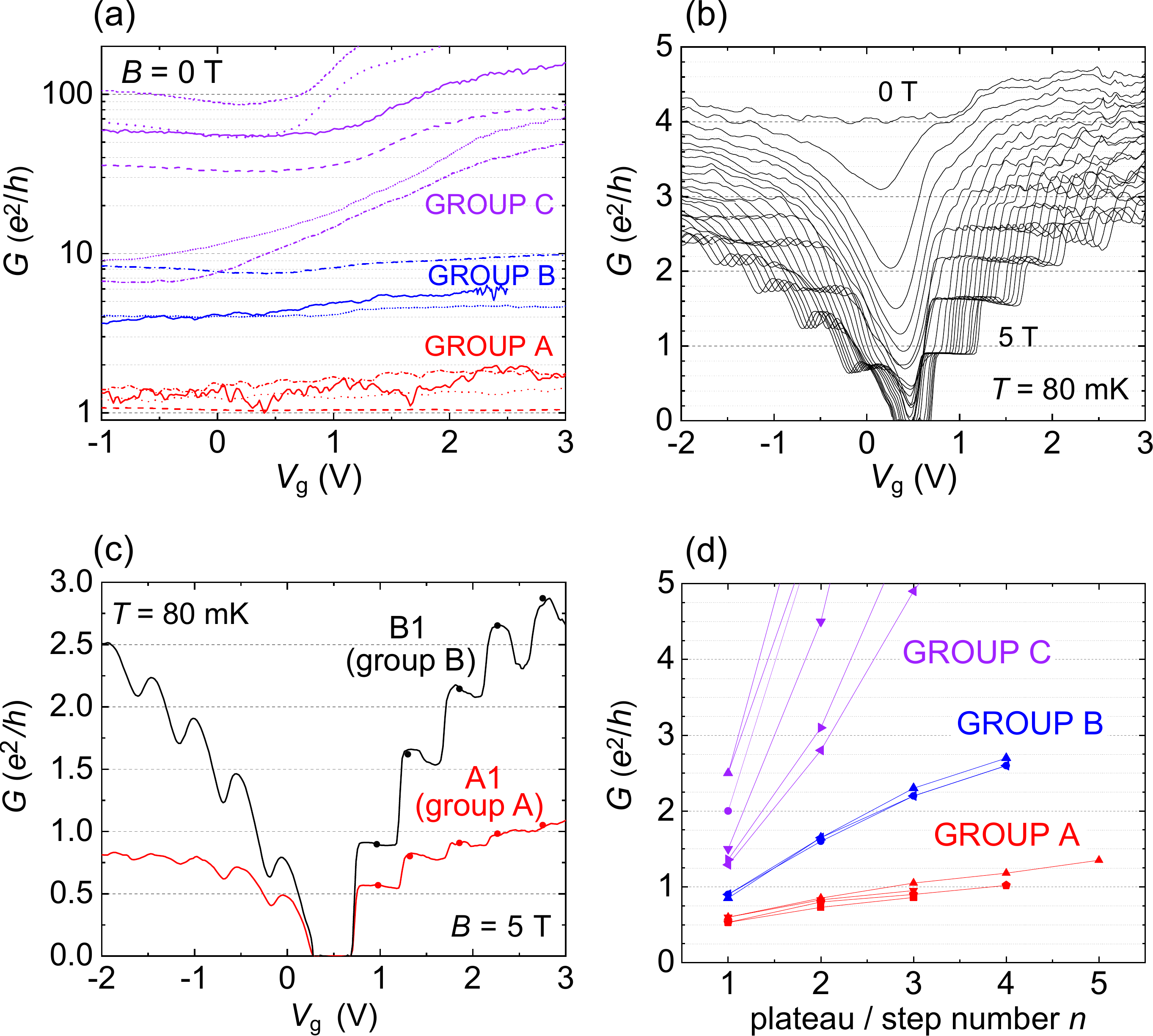}
\caption[Grouping] {
(a) The conductance $G(V_{\rm g})$  of the devices under study without any magnetic field. 
The observed $G(V_{\rm g})$ dependences are categorized into three distinct groups based on the conductance value: Group A (red, $G=1\ldots2$\,$e^2/h$), B (blue, $4\ldots10$\,$e^2/h$), and C (violet, $>30$\,$e^2/h$). While not all data are presented, each device analyzed unequivocally belongs to one of these groups, with none falling into the intermediate regions. The temperature ranges between \(T\) = 80 mK and 1.5 K for the different devices.
(b) $G(V_{\rm g})$ for device B1, belonging to group B, measured for magnetic field values ranging from $B = 0$ to $B=5$\,T  in increments of 200\,mT. 
(c) Comparison of $G(V_{\rm g})$ for devices A1 and B1, measured at $B = 5$\,T. 
(d) Dependence of the conductance step $G$ on the step number. A grouping at preferred values is observed for all devices from groups A and B. The temperature ranges between \(T\) = 80 mK and 1.5 K for the different devices.
}
\label{Figure3}
\end{figure}

We fabricated and investigated 15 QPCs with cross sections varying by more than one order of magnitude (see Appendix \ref{app:devices} for details). Consequently, the conductance of the devices at zero magnetic field, as shown in Fig.~\ref{Figure3}(a), ranges from 1 to 100\,$e^2/h$.  
Surprisingly, the measured $G(V_{\rm g})$ dependences are not uniformly distributed across the conductance range but fall into three distinct groups (A, B, and C) based on their conductance values. Each examined device can be assigned to one of these groups, with none falling into the gaps between them. Although the grouping might not be convincing at zero magnetic field, it becomes evident and well distinguished when a magnetic field is applied.
%While the grouping at zero magnetic field may seem unconvincing, the presence and distinction of the groups become clear in a magnetic field. 
Figure \ref{Figure3} (b) shows the dependence of $G(V_{\rm g})$ in a magnetic field ranging from 0 to 5\,T for device B1 from group B.  What distinguishes group B QPCs from group A is the sequence of step heights. Figure \ref{Figure3} (c) compares the step sequence of a representative device from group A, A1, with that of a representative from group B, B1.  The comparison of  $G(V_{\rm g})$ at $B=5$\,T, shows that the step heights for the B1 device are significantly higher than those for the A1 device. These distinct step sequences are representative of the behavior of all devices in groups A and B. 
Plotting the step height as a function of the step number, 
as shown in Fig.~\ref{Figure3} (d), reveals a clear grouping tendency: the difference in step height between samples in group A and group B is an order of magnitude greater than the difference within each group. 

Thus, all small and medium-sized QPCs do not exhibit a wide range of conductance values but rather fall into one of two groups, each with predetermined behavior in a magnetic field. Falling into either group A or B seems to correlate with the QPC cross-section. However, unlike the behavior of the conductance within each group, there is a notable uncertainty and overlap between the groups: samples in group A are characterized by cross-sections ranging from $5\times10^3$ to $1.8\times10^4$\,nm$^2$, while in group B the cross-section varies from $10^4$ to $2\times10^4$\,nm$^2$. The specific distribution of electrical charge within the QPC can act as a random and unpredictable factor influencing the tendency to belong to one of the groups. It is uncertain whether the observed grouping pattern results from the random distribution of a finite sample set or is governed by an underlying principle. Nevertheless, this ambiguity does not impact the subsequent analysis of the unusual step sequences. Notably, the models presented below are expected to account for samples that may, for instance, occupy intermediate positions between group A and group B, as shown in Figs.  \ref{Figure3} (a) and \ref{Figure3} (d).

Devices wider than 700 nm exhibit zero-field conductance above 30 $e^2/h$  and are categorized into group C. This group is distinguished not only by step heights greater than  $e^2/h$, but also by significant variation in this value [refer to Fig.~\ref{Figure3} (d)]. However, devices in this group possess such high conductance that they no longer adhere to the characteristics of QPCs; rather, they mark a transition to a macroscopic 2D system with a constriction. This transition is reflected in the altered form of $G(V_{\rm g})$, showcasing oscillations reminiscent of Shubnikov-de Haas phenomena rather than magnetic field-induced steps (not shown). While the study of this transition lies beyond the scope of this paper, our focus remains on samples from groups A and B exclusively. 

\begin{figure}
\centering
\includegraphics[width=1\columnwidth]{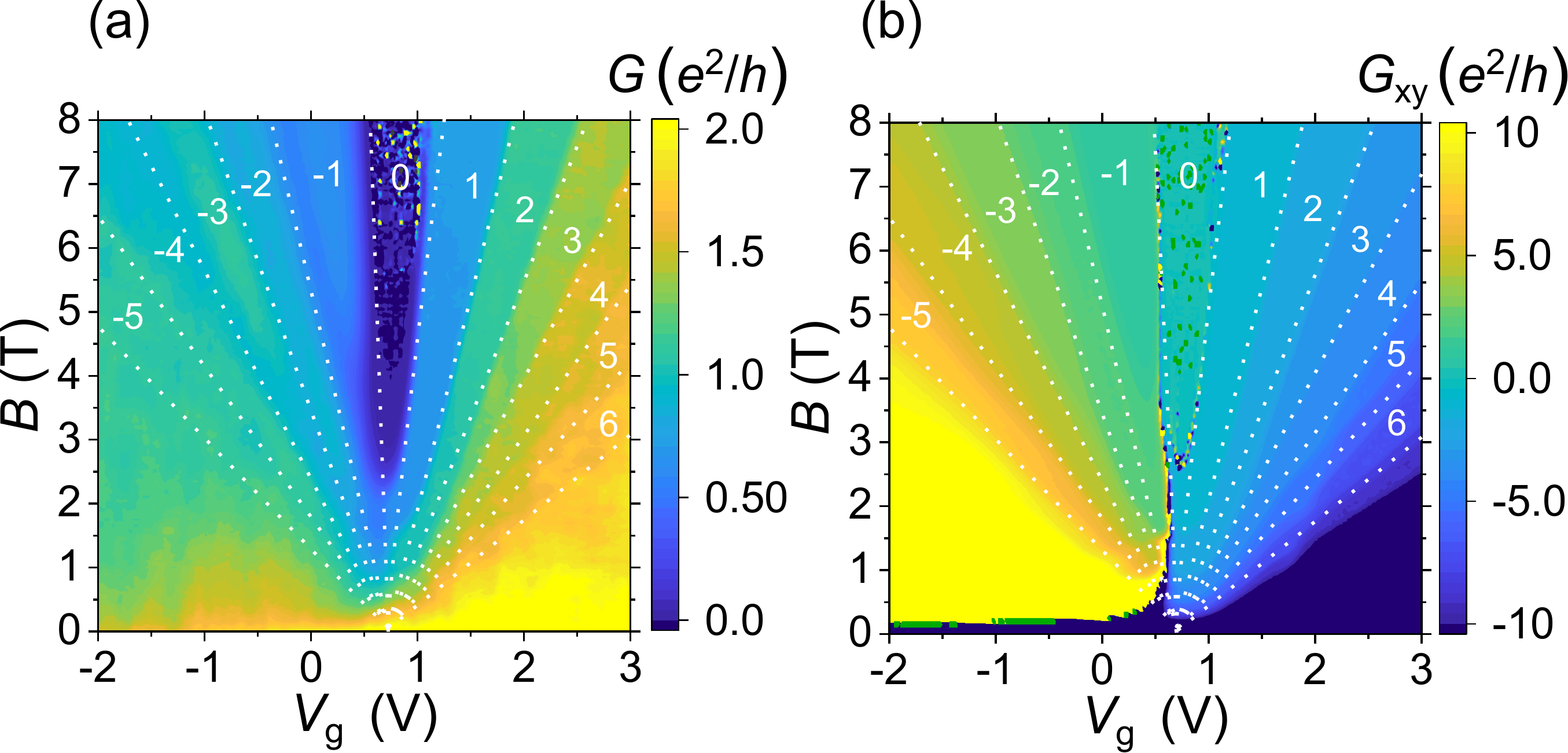}
\caption[] {
Color plots of (a) the longitudinal conductance of device A3 and (b) the macroscopic Hall conductance. Comparing the conductance across the QPC in (a) with the conductance of the macroscopic part in (b), we observe that transitions between steps occur at the same positions in both the QPC and the macroscopic part, determined by the filling factor of the bulk. This suggests that the QPC acts as a filter, allowing only a portion of the individual  edge states from the macroscopic part of the sample to pass through.}
\label{Figure4}
\end{figure}

Next, we analyze the correlation between the step positions in the conductance across the QPC and the quantum Hall steps outside the QPC as a function of $B$ and $V_g$ in Fig. ~\ref{Figure4} (a). The steplike behavior of $G(V_{\rm g},B)$ is evident from the abrupt color changes in Fig. \ref{Figure4} (a), with the step numbers indicated by the white labels. The differently colored segments form a fan that originates at the charge neutrality point. Although the conductance values differ, this fan closely matches the one shown in Fig. ~\ref{Figure4} (b), which visualizes the quantum Hall steps.  For $B>3$\,T, the Hall conductance exhibits quantized values given by $\nu e^2/h$, thus reflecting the filling factor $\nu$.   
This demonstrates that the step number in the QPC conductance reflects the number of filled Landau levels in the macroscopic regions outside the QPC, and, accordingly, the number of edge channels potentially flowing through the QPC. Since each edge QHE state has a conductance of $e^2/h$, and the height of the steps in $G$ is always smaller than $e^2/h$, it indicates that the QPC acts as a filter, partially transmitting the edge states from one side of the QPC to the other. Simultaneously, the transmission coefficient of this filter remains practically unchanged within one step, i.e., in the magnetic field range from 3 to 8\,T. 

In a first attempt to model the unusual sequence of plateaus, we resort to the Landauer-Büttiker (LB) formalism. In this formalism, the conductance of a multiterminal system in quantizing magnetic fields is expressed in terms of the transmission and reflection coefficients of the individual edge channels. The transmission coefficient $t$ of a QPC represents the number of one-dimensional states that successfully pass through it. For our case, we consider the six-terminal device shown in Fig.~\ref{Figure5} (a).  The contacts are numbered from 1 to 6, where contacts 1 and 4 serve as the source and drain, respectively, while the remaining contacts function as voltage probes.  For integer QHE filling factors, transport occurs along edge states, with the number of edge states denoted as $N$. Each edge state has a conductance of $e^2/h$. In one magnetic field orientation, these states propagate along the sample edges in a counterclockwise direction. Consequently, $N$ states flow from source 1 to contact 6. From contact 6, these $N$ states move toward the QPC, where they are split into $t$ states that pass through the QPC and $r$ states that are reflected toward contact 2, with the condition $N=t+r$. The deviation of the conductance plateaus from integer multiples of $e^2/h$, as in our experiments, indicates that $t$  (and consequently $r$) are not integers. Within the LB-formalism, the conductance across the QPC is given by $ G = \frac{e^2}{h} \frac{Nt}{N-t}$, from which we can derive $t = \frac{g \cdot N}{N + g}$, where $g = G/(e^2/h)$.  Using $N=G_{xy}/(e^2/h)$, we can derive $t$ from the experimental data. This is shown for sample A1 in Fig. \ref{Figure5} (b). The plot shows that the transmission at the first plateau is $\sim 0.4$. This means that of the single edge channel sketched in Fig. \ref{Figure1} (c),  40\% is transmitted and 60\% is reflected at the QPC. As the number of steps increases, the transmission is not proportional to the number of steps, but grows sublinearly.

\section{Theory: simulations}

\begin{figure}
\centering
\includegraphics[width=1\columnwidth]{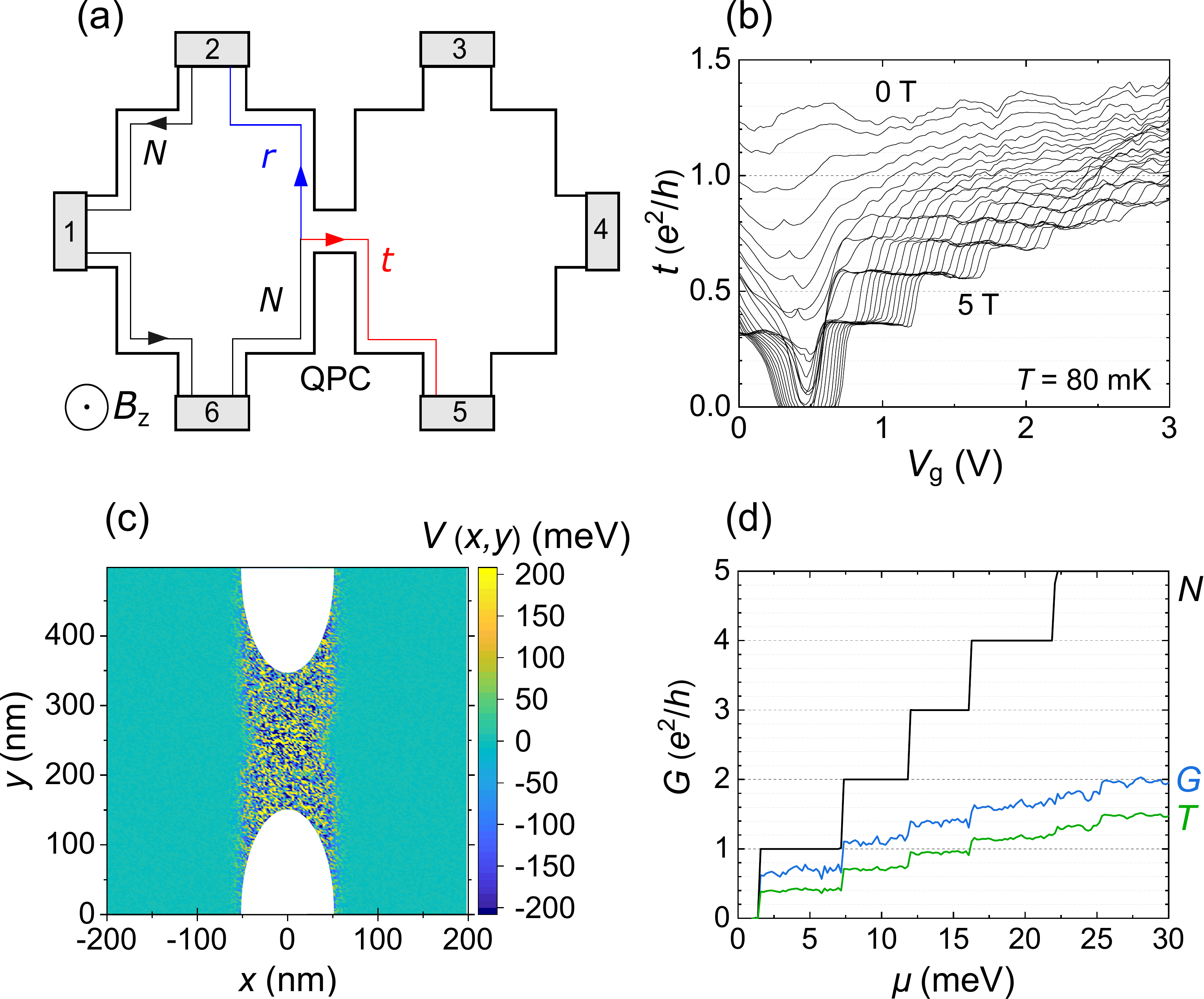}
\caption[] {
(a) Diagram of the multi-terminal device used to calculate the transmission coefficient $t$  in the Landauer-Büttiker formalism (not to scale). Out of $N$  edge states (depicted in black), $t$ modes (highlighted in red) are transmitted, while $r$ edge states (shown in blue) are reflected by the QPC. The mechanism is shown only on one side of the QPC (from left to right).
(b) The transmission coefficient $t(V_{\rm g})$ for device A1, calculated from the measured values of  $G(V_{\rm g})$ and $G_{\rm xy}(V_{\rm g})$ over the range $B = 0 - 5$\,T in steps of 200\,mT.
(c) An example of the electrostatic potential disorder distribution in the QPC region with the width $w = 200$\,nm, used for the conductance simulations.
(d) Numerically computed conductances employing the effective 2D model in Eq. \ref{eq:H_eff} for the Hall bar setup with $w = 200$\,nm shown in Fig. \ref{Figure5} (a) vs. chemical potential $\mu$ in a perpendicular magnetic field $B_{\rm z} = 5$\,T. The Hall conductance $G_{xy}$ (black) shows a spin non-degenerate step like structure with integer multiples of $e^2/h$ (denoted as $n$).  
}
\label{Figure5}
\end{figure}

In order to gain insight into the possible origins of the non-integer quantized conductance steps, we employ numerical tight-binding calculations with the Python package {\scshape{kwant}} \cite{Groth_2014}.
Note that a full 3D model is not tractable in transport simulations for the experimentally studied system dimensions, where a schematic illustration is shown in Fig. \ref{Figure5} (a).
As the experiment exhibits signatures of the 2D  quantum Hall effect in strong magnetic fields, we employ an effective model which describes a 2D electron conduction band.
The Hamiltonian which we consider is given by
\begin{multline}\label{eq:H_eff}
	H = \frac{\hbar^2}{2m_e}\left(k_x^2 + k_y^2-\mu+ V(x,y)\right) \sigma_0 \\ 
    + \alpha_R (k_x\sigma_y-k_y\sigma_x)+\frac{1}{2}g\mu_B B_z,
\end{multline}

where we include a Rashba-type spin-orbit coupling term, a Zeeman term and a spatially dependent scattering potential, $V$.
For all of the shown calculations, we fixed $m_e=0.06m_0$ with $m_0$ being the free electron mass, $\alpha_R=-0.015~\mathrm{meVnm}$ and $g=22.7$~\cite{neverov2020effective,König_2008,Essert_2016}.
The Hamiltonian $H$ is then discretized via the finite difference method on a discrete square lattice with width $w$ and length $L$.
The lattice constant is fixed to $a=2~\mathrm{nm}$ throughout this work.
Next, we attach four vertical and two horizontal leads to construct a six-terminal setup, as shown in Fig. \ref{Figure5} (a).
The latter have the same width as the central scattering region, while the former are separated from the horizontal leads by the distance $d_v$ and have a width $w_v$.
Additionally, we include the perpendicular magnetic field $B_z$ via standard Peierl’s substitution~\cite{peierls1933theory}, where we fix the gauge to $\mathbf{A} = (0,B_z x,0)$. 
Finally, we define the etched QPC region by defining elliptic areas stretching from the edges in the central part of the scattering region.
Additionally, we assume that the etching procedure not only shapes the device but also affects the potential landscape of the QPC zone.
For simplicity and to be as general as possible, we include such an effect using white noise random on-site potential $V(x,y)$.
This potential is added to the Hamiltonian and enters the tight-binding calculation as an on-site term.
We provide more details concerning the QPC area and the assumed scattering potential in  Appendix \ref{app:QPC_area_pot}.
%An representative  disorder set is shown in Fig.~\ref{Figure5}~(c).
We also implemented a more system-specific model for HgTe quantum wells~\cite{science-2006-BHZ} which would allow to study topological states in a bulk band gap, see Appendix \ref{app:2D BHZ simulations}. However, since no traces of such states were observed in experiment and both models give qualitatively similar results, we focus on the simpler system given by Eq. (\ref{eq:H_eff}).

We calculate the conductance components by numerically solving  the equation
\begin{equation}\label{eq:I_gV}
\mathbf{I} = \langle\hat{\mathbf{G}}\rangle_D \mathbf{V},
\end{equation}
where $\mathbf{I}$ and $\mathbf{V}$ are vectors containing all currents and voltages in the setup.
We define leads 4 and 5 as current probes, such that $\mathbf{I}=(0,0,0,0,I,-I)$ and set $V_5=0$.
Furthermore, $\langle.\rangle_D$ denotes an averaging over $D$ disorder sets. 
It is important to average first the full conductance matrix and solve then Eq. (\ref{eq:I_gV}).
Otherwise, by first solving the equation for each disorder set and averaging afterwards, the extracted longitudinal conductance leads to numerical divergences for small numbers of edge states.
The reason for that lies in the relation $G_{I, U} = G_{45, 32} = G_{xx} = \frac{e^2}{h} \frac{t}{N-t}$~\cite{Ferry_Goodnick_1997} (for conduction band edge states), where $t$ corresponds to the number of edge states which are transmitted through the QPC.
Experimentally, $t$ can be determined by measuring the diagonal conductance given by $G_d=\frac{e^2}{h}t$.
We see that whenever $t$ is close to $N$, one obtains a very small denominator and $G_{xx}$ diverges, and an adequate number of disorder averages cannot remove those contributions.

\section{Numerical modelling}

For the numerical simulations, we first fix the following geometric system parameters that are valid for all the shown results.
We will consider a setup with a width of $w=500~\mathrm{nm}$ and a length of $L=1600~\mathrm{nm}$.
Then, we attach vertical leads of width $w_v=250~\mathrm{nm}$ and $d_v=200~\mathrm{nm}$.
Next, we define the QPC region by fixing $\xi_x=50~\mathrm{nm}$ and $\xi_y=150~\mathrm{nm}$ such that we obtain a QPC width of $w_{\rm QPC} = 200~\mathrm{nm}$.
We set the magnetic field strength  to $B_z = 5~\mathrm{T}$, which results in a magnetic length of $l_B=\frac{\hbar}{eB_z}=11.47~\mathrm{nm}$.
Therefore, we expect the emergence of quantum Hall states in the QPC area.
For the etching-induced disorder we set $\xi_{bd}=110~\mathrm{nm}$, $s_x = 0.1$, $s_y=0.05$, and $K_{edge}=0.95~t$ with $t=\hbar^2/(2m_e)$.
The bulk disorder contribution is defined by setting $K_{bulk}=0.002~t$.
See Fig.~\ref{Figure5}~(c) for an example disorder profile.
Finally, we perform disorder averaging over 200 disorder sets to avoid conductance signatures related to an individual disorder potential.

We illustrate the results of our transport simulations in Fig.~\ref{Figure5} (d), where we plot different conductance components vs.~the chemical potential.
We show the absolute value of transversal Hall conductance, $|G_{xy}| = |G_{45,13}|$, in black, the longitudinal conductance, $G_{xx}=G_{45,01}$, in blue, and the diagonal conductance, $G_d=G_{45,03}$, in green.
The results exhibit well-quantized quantum Hall plateaus in $|G_{xy}|$ with multiples of $\frac{e^2}{h}$.
The lifting of the spin degeneracy stems from the spin-orbit coupling, and the included Zeeman effect.
Concerning the longitudinal conductance, $G_{xx}$, we observe plateau-like structures, where the conductance is quantized in non-integer values of $\frac{e^2}{h}$.
Furthermore, the increase in conductance from step to step is not given by a fixed constant.
Small remaining fluctuations  reflect the finite number of 200 disorder averages, which would be reduced further by including more disorder sets.
Still, the scattering of edge states in the energy range between two Landau levels appears to be independent of the chemical potential (gate voltage) in the numerical data.
Analogously, the diagonal conductance shows step-like features, and as mentioned above, it is directly proportional to the number of transmitted edge states $t$. 
Therefore, our results demonstrate that, indeed, transmission is increasing in steps.
Summarizing, we see that disorder in the QPC area can lead to clear step-like conductance features with non-integer quantization in terms of $e^2/ h$.

\vspace{1cm}

\section{Toy Model and Discussion}

\begin{figure}
\centering
\includegraphics[width=1\columnwidth]{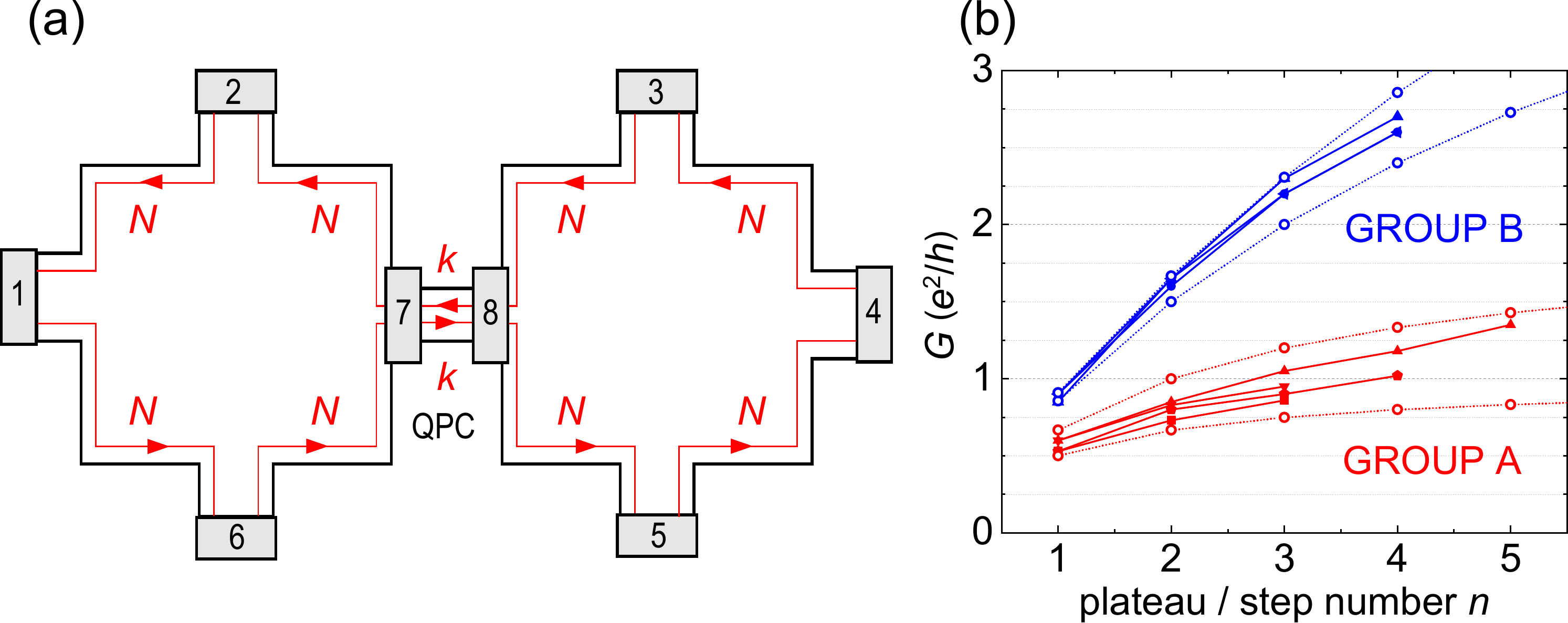}
\caption[] {(a) Landauer-Büttiker model including the introduction of two additional thermalization points at the entrance and the exit of the QPC (contacts 7 and 8). (b) Histogram of the step G on the step number [filled symbols, same as in Fig.~\ref{Figure3} (d)] for all devices supplemented by a calculated conductance $G = \frac{N \cdot k}{N + k}\frac{e^2}{h}$ (empty symbols) with $k = 1$ and $k=2$ (for group A) and $k = 6$ and $k = 10$ (for group B). All experimental data fall between the calculated dependences for each of the groups. }
\label{Figure6}
\end{figure}

The above numerical simulations show that steps with non-integer conductance values can be observed at a certain level of disorder, consistent with experimental data. However, despite the appeal of these calculations, the disorder amplitude remains merely an arbitrary fitting parameter. Since disorder in the junction appears to be the dominant factor, we adopted a modified Landauer-Büttiker-type toy model. In this model, we assume that the QPC input and output are relevant scattering centers, while the conductivity remains largely unaffected by scattering or dissipation outside these regions. In the Landauer-Büttiker model, additional scattering is represented by introducing two virtual 'contacts' near the QPC (contacts 7 and 8 in Fig.~\ref{Figure6} (a)). Including these virtual contacts leads to several consequences. 
It enables state mixing and thermalization within the contacts and bypasses the $N=t+r$ constraint from charge conservation. As a result,  $t$ and $r$ are replaced by a single coefficient $k$, determining the number of modes in the QPC, and thus its conductance, while charge conservation remains intact. 
The four-terminal conductance obtained for the given device geometry is defined by the formula 
\begin{equation}
\label{eq:LB}
    G = \frac{N \cdot k}{N + k}\frac{e^2}{h}
 \end{equation}
with $t = N \cdot k /(N + 2k)$ the transmission coefficient. In principle, $k$ can take any value independent of $N$. However, when using a fixed value of $k$ close to the value of the conductivity $G$ of the QPC at zero magnetic field [Fig.~\ref{Figure3} (a)], surprisingly, the results obtained from this
Landauer-Büttiker type of model are in good agreement with the experiment. Figure \ref{Figure6} (b) shows that all experimental data for group A lie between the two calculated curves obtained for $k = 1$ and $k=2$. Similarly, the data for group B falls within the calculated dependencies for  $k = 6$ and $k=10$. These values correspond to the minimum and maximum conductivity, respectively, in units of $e^2/h$ for all QPCs in each group. This finding suggests a complementary perspective: the system can be viewed as a series connection of two quantum Hall conductors with a magnetic field-independent resistor (i.e., the point contact) in between, which has a conductance of $k\cdot e^2/h$. In the Landauer-Büttiker model, the  resistance $R_{27}$ between contacts 2 and 7 at a plateau corresponds to the quantum Hall resistance $\frac{h}{e^2} \frac{1}{N}$, $R_{38}=0$ (noting that $R_{27}$ and $R_{38}$ interchange when the magnetic field is reversed), and $R_{78}=\frac{h}{e^2} \frac{1}{k}$, representing the zero-field resistance of the QPC. Note that $k$ can be non-integer. When summing these resistances to obtain the total resistance across the point contact and converting to conductance, we get  $G = \frac{N \cdot k}{N + k}\frac{e^2}{h}$, which matches Eq.~(\ref{eq:LB}). 
This perspective suggests that arbitrary plateau sequences can be generated as long as the resistance of the point contacts is (nearly) independent of the applied magnetic field. While this view overlooks the energy bands in the 3D topological insulator-based QPCs and the material's topological nature, it accurately captures the most prominent features. 

\vspace{-10pt}
\section{Conclusion}

In this work, we have shown that quantum point contacts based on HgTe 3D topological insulators exhibit quantization of the conductance at non-integer values when subjected to a magnetic field.
The occurrence of steps can be reproduced by numerical tight-binding calculations using the Python package {\scshape{kwant}}. The calculations suggest that disorder in the point contact plays the leading role in the atypical quantization sequences. Modeling such disorder phenomenologically using a LB model, we find that the conductance steps can be described by 
Eq.~(\ref{eq:LB}) 
where $N$ is the number of edge modes (filling factor) in the wide regions outside the QPC and $k$ is the 
conductance in units of $e^2/h$ at zero magnetic field. 

%We love QPCs :)
\begin{acknowledgements}
The work was supported by the European
Research Council (ERC) under the European Union's Horizon
2020 research and innovation program (Grant Agreement No.
787515, "ProMotion") and by the
Deutsche Forschungsgemeinschaft (DFG, German Research Foundation) within Project-ID 314695032 – SFB 1277.
\end{acknowledgements}

\appendix

\section{List of devices}
\label{app:devices}

\begin{table}[h!]
\centering % used for centering table
\begin{tabular}{|c| c| c| c| c|} % centered columns (4 columns)
 \hline
Device      & \(w\) (nm)    & \(h\) (nm)    &  \(A\) (nm²) & First step in\\ & & & & $G$ $(e^2/h)$ \\
\hline  
{A1}       &  580      &  30       & 17 400    & 0.6\\
\hline
{A2}        & 140       & 50        & 7000     & 0.6\\
\hline
A3        & 95        & 50        & 4750     & 0.55\\
\hline
A4        & 85        & 50        & 4250     & 0.55\\
\hline
{B1}       &  710      &  30       & 21 300    & 0.9\\
\hline
{B2}        &  200      & 50        & 10 000 & 0.87\\     
\hline
{B3}       & 135       & 80        & 10 800 & 0.85\\
\hline
{C1}       &  580     & 50        & 29 000 & 1.3 \\
\hline
{C2}       &  580     & 50        & 29 000 & 1.3 \\
\hline
{C3}       &  1,000     & 50        & 50 000 & 2.0 \\
\hline
{C4}       & 2,000      & 50 &        100 000 & 2.0\\  
%\hline
%{C5}       & 5,000      & 80 &        400,000 & 2.7\\
%\hline
%{C6}       & 10,000      & 80 &        800,000 & 3.0\\
% [1ex] adds vertical space
\hline %inserts single line
\end{tabular}
    \caption{List of devices, including their width and height of the HgTe layer, cross section \(A\) and the value of the first step in the conductance.}
    \label{devices}
\end{table}

\section{Details about QPC area and potential}
\label{app:QPC_area_pot}

As already mentioned in the main text, we define the etched QPC region by defining elliptic areas stretching from the edges in the central part of the scattering region. 
This is done by deleting sites which fulfill the conditions 
\begin{equation}\label{eq:elliptic_edge}
\frac{(x-L/2)^2}{(\xi_x)^2}+\frac{y^2}{(\xi_y)^2} < 1,
\end{equation}

and
\begin{equation}
\frac{(x-L/2)^2}{(\xi_x)^2}+\frac{(y-w)^2}{(\xi_y)^2} < 1.
\end{equation}

\begin{figure}
\centering
\includegraphics[width=\columnwidth]{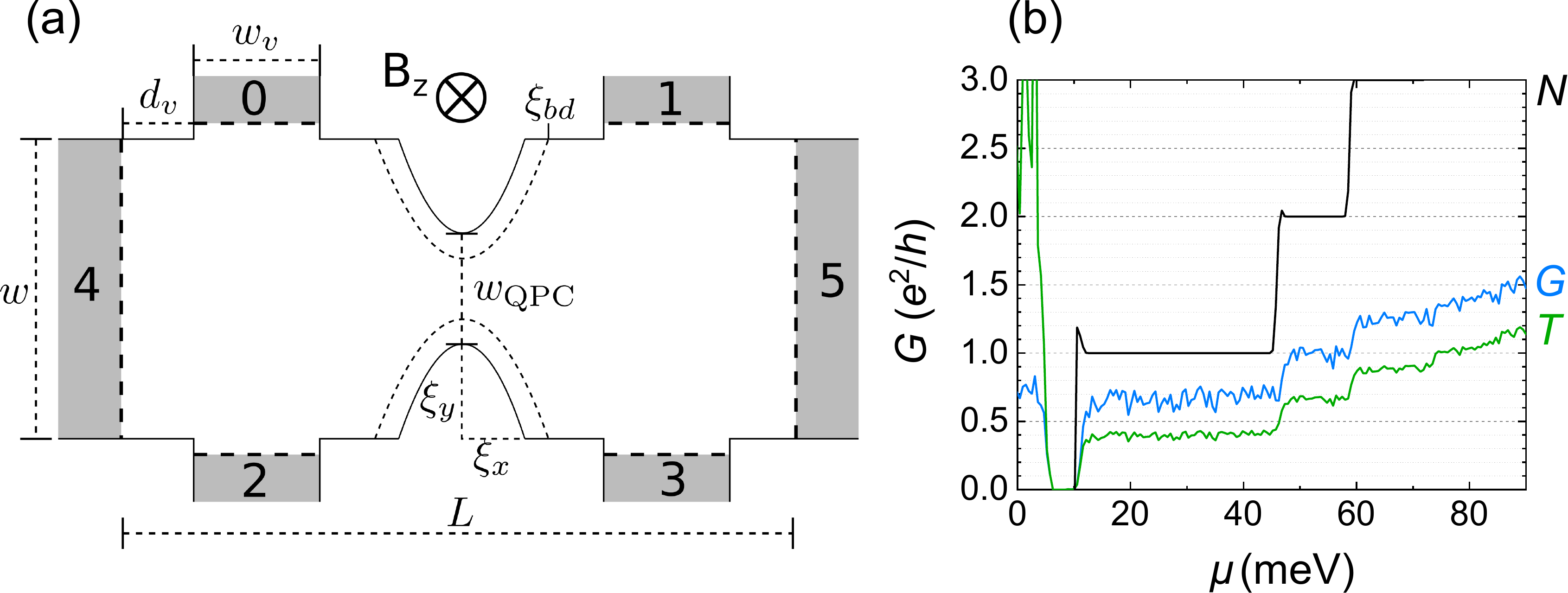}
\caption[] {(a) Hall bar schematic considered in the tight-binding simulations illustrating the important length and geometric parameters.
(b) Numerically computed conductances for the Hall bar setup with $w = 200$\,nm shown in Fig. \ref{Figure5}(a) vs. chemical potential $\mu$ for a topological choice of parameters in a perpendicular magnetic field $B_{\rm z} = 5$\,T, based on the 2D BHZ model. The Hall conductance $G_{xy}$ (black) shows a spin non-degenerate step like structure with integer multiples of $e^2/h$ (denoted as $n$). 
} 
\label{fig:system_schematic}
\end{figure}

The parameters $\xi_x$ and $\xi_y$ tune the spatial extent of the etched regions and are illustrated in Fig.~\ref{fig:system_schematic}~(a).
The width of the central QPC region is then given by $w_{\rm QPC} = w - 2 \xi_y$.
However, for experimentally studied QPC widths, the coupling induced by the pure geometric confinement between the emerging edge states is not sufficient to show any scattering and therefore a finite longitudinal resistance.
Therefore, we assume that the etching procedure is not only shaping the device, but also strongly influences the potential in the QPC.
We want to point out that we allow for the disorder to consist of two distinct parts.
On the one hand, $V_{bulk}(x,y)$ describes the disorder which is already present in the material before the etching procedure and its magnitude is tuned by the value $K_{bulk}$.
This component is probably very small, as the studied mesas exhibit very long phase coherence lengths~\cite{Ziegler_Phd}.
The second component, namely $V_{edge}(x,y)$ describes the disorder which is induced by the sample production and the strength is tuned by $K_{edge}$.
We want to highlight that the above choice of the disorder potential is only done by assuming a smooth scaling of white noise when approaching the etched areas.
Still, it suffices to show that simple on-site disorder can lead to the formation of robust steps in the longitudinal conductance.
The potential $V(x,y)$ which we used to calculate the results of the main text is defined as

\begin{equation}
	V(x,y) = V_{edge}(x,y)  f_y(x,y) f_x(x)  +   V_{bulk}(x,y),
\end{equation}

with
\begin{align}
	f_y(x,y) &= 0.5\{-\tanh[s_y(y-e(x,y)-\xi_{bd})] + \nonumber \\
	\phantom{f_y(x,y) }&\phantom{= 0.5\{-t }\tanh[s_y(y+e(x,y)-w_x+\xi_{bd})] \} +1,\\
	f_x(x)   &= 0.5\{\tanh[s_x(x-L/2+\xi_x)] - \nonumber \\
	\phantom{f_x(x)   }&\phantom{= 0.5\{ }\hspace{1.5pt} \tanh[s_x(x-L/2-\xi_x)]\},
\end{align}

and
\begin{align}
V_{edge}(x,y) &= K_{edge} n(x,y) \\ 
V_{bulk}(x,y) &= K_{bulk} n(x,y).
\end{align}

The parameters $s_x$ and $s_y$ tune the smoothness of the potential scaling, while $\xi_{bd}$ defines something like the “penetration depth” of the etching-induced disorder close to the boundaries.
Additionally, $e(x,y)$ is an offset determined by the edge parametrization given by Eq.~\ref{eq:elliptic_edge}. 
Finally, $n(x,y)$ corresponds to random numbers which are drawn from a uniform distribution in $[0,1)$.

\section{Further simulations: 2D BHZ model}
\label{app:2D BHZ simulations}

The so-called 2D Bernevig-Hughes-Zhang (BHZ) Hamiltonian~\cite{science-2006-BHZ} reads in the basis $|E1\uparrow\rangle$, $|H1\uparrow\rangle$, $|E1\downarrow\rangle$, and $|H1\downarrow\rangle$
\begin{equation}
H_ {BHZ} = \left(
\begin{tabular}{cccc}
\multicolumn{2}{c}{\multirow{2}{*}{$h(\mathbf{k})$}} & 0 & 0 \\
\multicolumn{2}{c}{} & 0 & 0                            \\
0 & 0 & \multicolumn{2}{c}{\multirow{2}{*}{$h^*(-\mathbf{k})$}}\\
0 & 0 & \multicolumn{2}{c}{} 
\end{tabular}    \right),
\end{equation}

with
\begin{equation}
h(\mathbf{k}) = 
\begin{pmatrix}
C - (B + D) \mathbf{k}^2 + M & A k_+\\
A k_- & C + (B - D) \mathbf{k}^2 - M
\end{pmatrix}.
\end{equation}

The momentum operators are given by $\mathbf{k}^2 = k_x^2 + k_y^2$ and $k_\pm = k_x \pm i k_y$. 
We use the Hamiltonian parameters $A=364.5~\mathrm{meV}\cdot\mathrm{nm}$, $B=-686~\mathrm{meV}\cdot\mathrm{nm}^2$, \(C=0\), $D=-512~\mathrm{meV}\cdot\mathrm{nm}^2$, and $M=-10~\mathrm{meV}$ which correspond to a 7 nm thick HgTe quantum well~\cite{Rothe_2010,Essert_2016}.
This model allows us to examine a system closely related to the experimental samples with valence and conduction band as well as topological states in a bulk band gap.
Furthermore, we also take into account additional Hamiltonian terms originating from bulk inversion asymmetry $H_{BIA}$ and structure inversion $H_{SIA}$, as well as the influence of the Zeeman effect $H_Z$.
We also take into account additional Hamiltonian terms originating from bulk inversion asymmetry and structure inversion asymmetry~\cite{Rothe_2010,König_2008,Essert_2016}.
Those two contributions are given by
\begin{equation}
H_{BIA} = 
\begin{pmatrix}
0 & 0 & 0 & -\Delta \\
0 & 0 & \Delta & 0  \\
0 & \Delta & 0 & 0  \\
-\Delta & 0 & 0 & 0 \\
\end{pmatrix}
\end{equation}

and 
\begin{equation}
H_{SIA} = 
\begin{pmatrix}
0 & 0 & -iRk_- & 0 \\
0 & 0 & 0 & 0  \\
iRk_+ & 0 & 0 & 0  \\
0 & 0 & 0 & 0 \\
\end{pmatrix},
\end{equation}

respectively, with the parameters $\Delta = 1.6~\mathrm{meV}$ and $R = -15.6~\mathrm{meV}\cdot\mathrm{nm}$.
For the latter term we simplified the Hamiltonian and neglect the linear dependence of $R$ on the gate voltage induced electric field.
Furthermore, we include the Zeeman effect via~\cite{König_2008}
\begin{equation}
H_Z = 
\begin{pmatrix}
g_E\mu_B B_z & 0 & 0 & 0 \\
0 & g_H\mu_B B_z & 0 & 0  \\
0 & 0 & -g_E\mu_B B_z & 0  \\
0 & 0 & 0 & -g_H\mu_B B_z \\
\end{pmatrix},
\end{equation}

with the g-factors given by $g_E = 22.7$ and $g_H = -1.21$.

Note that we fix the same geometric parameters as in the main text for the simulation of this Hamiltonian.
However, we choose for the disorder components $K_{edge}=750~\mathrm{meV}$ and $K_{bulk}=30.3~\mathrm{meV}$ to match the different energy scale.

Results are shown in Fig.~\ref{fig:system_schematic} (b), where we consider a HgTe sample in the topological $M/B>0$ regime and again plot different conductance components vs. the chemical potential.

\bibliography{literature}

\end{document}